\documentclass[aps,onecolumn,a4paper,oneside,preprint,tightenlines,%
nobibnotes,nofootinbib,amsfonts,amssymb,amsmath,eqsecnum,showpacs,%
showkeys]{revtex4}

\newcommand{\cN}{\mathcal{N}}
\newcommand{\cV}{\mathcal{V}}
\newcommand{\cW}{\mathcal{W}}
\newcommand{\bH}{\boldsymbol{H}}
\newcommand{\bQ}{\boldsymbol{Q}}
\newcommand{\tH}{\tilde{H}}
\newcommand{\tP}{\tilde{P}}
\newcommand{\tcV}{\tilde{\cV}}
\newcommand{\tw}{\tilde{w}}
\newcommand{\tvph}{\tilde{\varphi}}

\newcommand{\bbN}{\mathbb{N}}

\newcommand{\bbR}{\mathbb{R}}
\newcommand{\bbC}{\mathbb{C}}
\newcommand{\rme}{\mathrm{e}}
\newcommand{\rmd}{\mathrm{d}}

\newcommand{\Llra}{\Longleftrightarrow}

\newcommand{\braket}[1]{\bigl\langle{#1}\bigr\rangle}
\newcommand{\rmT}{\mathrm{T}}
\newcommand{\fF}{\mathfrak{F}}
\newcommand{\gd}{\operatorname{gd}}
\newcommand{\ef}{\operatorname{Erf}}
\newcommand{\sech}{\operatorname{sech}}

\begin{document}


%
%

\title{Effect of Position-dependent Mass on Dynamical Breaking
of Type B and Type $X_{2}$ $\cN$-fold Supersymmetry}
\author{Bikashkali Midya}
\email{bikash.midya@gmail.com}
\author{Barnana Roy}
\email{barnana@isical.ac.in}
\affiliation{Physics and Applied Mathematics Unit, Indian Statistical
 Institute,\\ Kalkata 700108, India}
\author{Toshiaki Tanaka}
\email{toshiaki@post.kek.jp}
\affiliation{Department of Physics, National Cheng Kung University,\\
 Tainan 701, Taiwan, R.O.C.\\
 National Center for Theoretical Sciences, Taiwan, R.O.C.}
\thanks{Present address: Institute of Particle and Nuclear Studies, High Energy
 Accelerator Research Organization (KEK), 1-1 Oho, Tsukuba, Ibaraki 305-0801, Japan}


\begin{abstract}

We investigate effect of position-dependent mass profiles on dynamical
breaking of $\cN$-fold supersymmetry in several type B and type $X_{2}$
models. We find that $\cN$-fold supersymmetry in rational potentials in
the constant-mass background are steady against the variation of mass
profiles. On the other hand, some physically relevant mass profiles can
change the pattern of dynamical $\cN$-fold supersymmetry breaking in
trigonometric, hyperbolic, and exponential potentials of both type B
and type $X_{2}$. The latter results open the possibility of detecting
experimentally phase transition of $\cN$-fold as well as ordinary
supersymmetry at a realistic energy scale.

\end{abstract}





\preprint{TH-1536}

\maketitle

\section{Introduction}
\label{sec:intro}

In recent years, the study of quantum mechanical systems
with a position-dependent mass (PDM) have attracted a lot of interest due
to their relevance in describing the physics of many microstructures of
current interests, such as compositionally graded crystals~\cite{GK93},
semiconductor heterostructure~\cite{MB84}, quantum dots~\cite{SL97}, $^3$He
clusters~\cite{BPGHN97}, metal clusters~\cite{PSC94} etc. The concept of
PDM comes from the effective-mass approximation~\cite{Wa37,Sl49} which is
a useful tool for studying the motion of carrier electrons in pure crystals
and also for the virtual-crystal approximation in the treatment of
homogeneous alloys (where the actual potential is approximated by a
periodic potential) as well as in graded mixed semiconductors (where
the potential is not periodic). Recent interest in this field stems from
extraordinary development in crystal-growth
techniques like molecular beam epitaxy, which allow the production of
nonuniform semiconductor specimen with abrupt heterojunctions~\cite{Ba88}.
In these mesoscopic materials, the effective mass of the charge carrier
are position dependent. Consequently, the study of the position-dependent
mass Schr\"{o}dinger equation (PDMSE) becomes relevant for deeper
understanding of the non-trivial quantum effects observed on these
nanostructures. It has also been found that such equations appear in
many different areas. For example, it has been shown that constant mass
Schr\"{o}dinger equations in curved space and those based on deformed
commutation relations can be interpreted in terms of PDMSE~\cite{QT04}.
The PDM also appear in nonlinear oscillator~\cite{CRS07,MR09b} and
$\mathcal{PT}$-symmetric cubic anharmonic oscillator~\cite{Mo05b}.
The most general form of the PDM Hamiltonian proposed by von
Roos~\cite{Ro83} is defined by
\begin{align}
H = -\frac{1}{4}\left(m(q)^{\alpha}\frac{\rmd}{\rmd q}m(q)^{\beta}
 \frac{\rmd}{\rmd q}m(q)^{\gamma}+m(q)^{\gamma}\frac{\rmd}{\rmd q}
 m(q)^{\beta}\frac{\rmd}{\rmd q}m(q)^{\alpha}\right) + V(q),
\end{align}
where the ambiguity parameters $\alpha$, $\beta$, $\gamma$ are related by
$\alpha+\beta+\gamma=-1$. The above Hamiltonian always has the following form:
\begin{align}
H=-\frac{1}{2m(q)}\frac{\rmd^{2}}{\rmd q^{2}} + \frac{m'(q)}{2m(q)^{2}}
 \frac{\rmd}{\rmd q} + U(q),
\end{align}
where the effective potential $U(q)$ is given by
\begin{align}
U(q) = V(q) -(\alpha+\gamma) \frac{m''(q)}{4 m(q)^2} + (\alpha\gamma
 + \alpha+\gamma) \frac{m'(q)^2}{2m(q)^3}.
\end{align}

It is quite natural that physical interests just described above have
also enhanced the studies on exact solutions to PDMSE \cite{DCH98,MI99,PRCGP99,%
DA00,RR02,GOGU02,SL03,RR01,Al02,GN07,MRR10,KKS05,Qu06,BGQR05,MR09a,Ch04,RR05}
by employing various methods e.g. supersymmetric (SUSY) quantum
mechanics~\cite{Wi81} and point canonical transformation~\cite{KO90}
to mention a few. Later, PDM quantum systems were successfully
formulated in the framework of $\cN$-fold SUSY in Ref.~\cite{Ta06a},
which has provided until now the most general tool for constructing
a PDM system which admits exact solutions because of its equivalence to weak
quasi-solvability. To avoid confusion, we here note that $\cN$-fold SUSY is different
from \emph{nonlinear SUSY} which has been long employed since the work
by Samuel and Wess~\cite{SW83} in 1983 to indicate the nonlinearly realized SUSY
originated from the work by Akulov and Volkov~\cite{VA72} in 1972.
For a review of $\cN$-fold SUSY see Ref.~\cite{Ta09}, while for recent works on
nonlinear SUSY see, e.g., Ref.~\cite{ST10} and references cited therein.

Very recently, new classes of exactly solvable PDM quantum systems whose
eigenfunctions are expressible in terms of so-called $X_{1}$ polynomials
were constructed in Ref.~\cite{MR09a}. The new findings of $X_{n}$
polynomials ($n\geq1$) were associated with the more fundamental
mathematical concept of exceptional polynomial subspaces of codimension $n$
introduced in Refs.~\cite{GKM07,GKM08b,GKM08a}, whose origin can be traced
back to the pioneering work on the classification of monomial spaces
preserved by second-order linear differential operators~\cite{PT95}.

The purpose of the present paper is two-fold. The first one is to bring
the purely mathematical concept of exceptional polynomial subspaces into
more physical settings by allowing the position dependence of mass
(in a spirit similar to Ref.~\cite{MR09a}) in the framework of $\cN$-fold
SUSY. In the constant-mass case, form of potentials related to exceptional
polynomial systems is very limited. Thus, we can enlarge the physical
applicability of the mathematical concept by introducing PDM to quantum
systems. On the other hand, the framework of $\cN$-fold SUSY enables us to
talk about the physical phenomenon of dynamical $\cN$-fold SUSY breaking.
The second purpose is actually to examine effect of PDM profiles on
dynamical breaking of $\cN$-fold SUSY.
In this respect, it is rather surprising that there have been few papers,
like Ref.~\cite{MR09b}, where broken as well as unbroken SUSY is described
in PDM backgrounds depending on the mass profiles. One of the main reasons
would be that SUSY has been mostly used just as a technique to obtain exact
solutions.
The true significance of the Witten's SUSY quantum mechanics~\cite{Wi81},
however, rather resides in the nonperturbative aspects of dynamical SUSY
breaking. Hence, one of our main purposes is, in other words, to examine
change of nonperturbative nature of quantum systems caused by variations
of mass profiles in view of dynamical $\cN$-fold SUSY breaking.

The paper is organized as follows. In Section~\ref{sec:NSPDM}, we provide
a self-contained review of $\cN$-fold SUSY in a PDM background, especially
for those who are not familiar with the subject. We also summarize
mathematical structure of type B and type $X_{2}$ $\cN$-fold SUSY.
In Section~\ref{s3}, we construct several $\cN$-fold SUSY PDM quantum
systems and examine dynamical $\cN$-fold SUSY breaking in different PDM
backgrounds. The first three models of type B $\cN$-fold SUSY have rational,
trigonometric, and exponential potentials in the constant mass case. We
show in particular that the models whose bound state eigenfunctions were
shown to be expressed in terms of $X_{1}$ polynomials in Ref.~\cite{Qu08b}
for the constant mass case and in Ref.~\cite{MR09a} for the PDM cases can be
obtained as type B systems. The last three models of type $X_{2}$ $\cN$-fold
SUSY have rational, hyperbolic, and exponential potentials in the constant
mass case. For both types of $\cN$-fold SUSY, we find that the rational
potentials have steady $\cN$-fold SUSY against variation of mass profile
while all the other types of potentials can receive effect of PDM on their
dynamical breaking of $\cN$-fold SUSY. Finally, we summarize the results
and discuss their implications and prospects in Section~\ref{s5}.

\section{Review of $\cN$-fold Supersymmetry in a PDM background}
\label{sec:NSPDM}

An $\cN$-fold SUSY one-body quantum mechanical system with PDM is composed
of a pair of PDM Hamiltonians
\begin{align}
H^{\pm}=-\frac{1}{2m(q)}\frac{\rmd^{2}}{\rmd q^{2}}+\frac{m'(q)}{2m(q)^{2}}
 \frac{\rmd}{\rmd q}+U^{\pm}(q),
\label{eq:H+-}
\end{align}
and an $\cN$th-order linear differential operator
\begin{align}
P_{\cN}^{-}=m(q)^{-\cN/2}\frac{\rmd^{\cN}}{\rmd q^{\cN}}
 +\sum_{k=0}^{\cN-1}w_{k}^{[\cN]}(q)\frac{\rmd^{k}}{\rmd q^{k}},
\end{align}
which satisfy the following intertwining relations
\begin{align}
P_{\cN}^{-}H^{-}=H^{+}P_{\cN}^{-},\qquad P_{\cN}^{+}H^{+}=H^{-}P_{\cN}^{+}.
\label{eq:inter}
\end{align}
In the above, $P_{\cN}^{+}$ is the transposition~\cite{AS03} of $P_{\cN}^{-}$ given by
\begin{align}
P_{\cN}^{+}=(P_{\cN}^{-})^{\rmT}=\left(-\frac{\rmd}{\rmd q}
 \right)^{\cN}m(q)^{-\cN/2}+\sum_{k=0}^{\cN-1}\left(
 -\frac{\rmd}{\rmd q}\right)^{k}w_{k}^{[\cN]}(q).
\end{align}
Actually, the two relations in (\ref{eq:inter}) are not independent;
the first implies the second and vice versa, since the PDM
Hamiltonians (\ref{eq:H+-}) are invariant under the transposition
$(H^{\pm})^{\rmT}=H^{\pm}$.

One of the significant consequences of the intertwining relations
(\ref{eq:inter}) is \emph{weak quasi-sovability}, that is, $H^{\pm}$ preserves
a finite-dimensional linear space $\cV_{\cN}^{\pm}$ spanned by the kernel
of the operator $P_{\cN}^{\pm}$
\begin{align}
H^{\pm}\cV_{\cN}^{\pm}\subset\cV_{\cN}^{\pm},\qquad
 \cV_{\cN}^{\pm}=\ker P_{\cN}^{\pm}.
\label{eq:WQS1}
\end{align}
Each space $\cV_{\cN}^{\pm}$ is called a solvable sector of $H^{\pm}$.
Except for the $\cN=2$ case (cf., Refs.~\cite{Ta09,Ta03a}), virtually all the $\cN$-fold
SUSY systems so far found admit analytic expression of $\cV_{\cN}^{\pm}$ in closed
form, and thus are \emph{quasi-solvable}. In addition, it sometimes happens when
either $H^{-}$ or $H^{+}$ does not depend essentially on $\cN$ and preserves
an infinite flag of the solvable sectors
\begin{align}
\cV_{1}^{-/+}\subset\cV_{2}^{-/+}\subset\cdots\subset\cV_{\cN}^{-/+}
 \subset\cdots.
\end{align}
In this case, it is said to be \emph{solvable}, which is a necessary
condition for exact solvability. We note that $H^{-}$ and $H^{+}$ are
usually simultaneously solvable due to the intertwining relations
(\ref{eq:inter}).

A set of an $\cN$-fold SUSY system $H^{\pm}$ and $P_{\cN}^{\pm}$ provides
a representation of $\cN$-fold superalgebra defined by
\begin{align}
\bigl[\bQ_{\cN}^{\pm},\bH\bigr]=\bigl\{\bQ_{\cN}^{\pm},\bQ_{\cN}^{\pm}
 \bigr\}=0,\qquad\bigl\{\bQ_{\cN}^{-},\bQ_{\cN}^{+}\bigr\}=2^{\cN}
 \mathsf{P}_{\cN}(\bH),
\label{eq:NSalg}
\end{align}
where $\mathsf{P}_{\cN}(x)$ is a monic polynomial of degree $\cN$ in $x$.
Indeed, it is realized by defining $\bH$ and $\bQ_{\cN}^{\pm}$ as
\begin{align}
\bH=H^{-}\psi^{-}\psi^{+}+H^{+}\psi^{+}\psi^{-},\quad
 \bQ_{\cN}^{+}=P_{\cN}^{-}\psi^{+},\quad\bQ_{\cN}^{-}=P_{\cN}^{+}\psi^{-},
\end{align}
where $\psi^{\pm}$ is a pair of fermionic variables satisfying
$\{\psi^{\pm},\psi^{\pm}\}=0$ and $\{\psi^{-},\psi^{+}\}=1$. It is easy
to check that the above $\bH$ and $\bQ_{\cN}^{\pm}$ satisfy the first
part of algebra (\ref{eq:NSalg}). In particular, the intertwining
relations in (\ref{eq:inter}) guarantee the commutativity of $\bH$ and
$\bQ_{\cN}^{\pm}$. Regarding the second part of algebra, the monic
polynomial $\mathsf{P}_{\cN}$ is given, in the above representation,
by~\cite{AS03,Ta06a}
\begin{align}
\mathsf{P}_{\cN}(\bH)=\det\left(\bH-H^{\pm}\bigr|_{\cV_{\cN}^{\pm}}\right),
\end{align}
namely, the characteristic polynomial for $H^{\pm}$ restricted to the
solvable sectors $\cV_{\cN}^{\pm}$.

Whether $\cN$-fold SUSY of the system under consideration is dynamically
broken is determined by a property of the solvable sectors $\cV_{\cN}^{\pm}$
since they characterize $\cN$-fold SUSY states, namely, states annihilated
by the pair of $\cN$-fold supercharges $\bQ_{\cN}^{\pm}$. Let $|0\rangle$
and $|1\rangle$ be the fermionic vacuum and the one fermion state,
respectively, which satisfy
\begin{align}
\psi^{-}|0\rangle=0,\qquad |1\rangle=\psi^{+}|0\rangle.
\end{align}
Then, superstates $|\Psi_{0}^{-}\rangle=\Psi_{0}^{-}(q)|0\rangle$ and
$|\Psi_{0}^{+}\rangle=\Psi_{0}^{+}(q)|1\rangle$, respectively, are
annihilated by both of $\bQ_{\cN}^{\pm}$
\begin{align}
\bQ_{\cN}^{\pm}|\Psi_{0}^{-}\rangle=0,\qquad\bQ_{\cN}^{\pm}|\Psi_{0}^{+}
 \rangle=0,
\label{eq:NSsts}
\end{align}
if and only if $\Psi_{0}^{-}(q)\in\cV_{\cN}^{-}$ and $\Psi_{0}^{+}(q)\in
\cV_{\cN}^{+}$, respectively. However, such states do not necessarily
satisfy physical requirements. Suppose $S\subset\bbC$ is a domain
where both of the Hamiltonians $H^{\pm}$ have no singularities and are
thus naturally defined, and $\fF(S)$ is a linear space of complex functions
in which both of $H^{\pm}$ act. In a usual physical application, the domain
$S$ is the real line $\bbR$ or a real half-line $\bbR_{+}=(0,\infty)$,
and the linear space $\fF$ is a Hilbert space $L^{2}$, so that
$\fF(S)=L^{2}(\bbR)$, or $L^{2}(\bbR_{+})$. In the latter cases, the physical
requirement is the normalizability (square integrability) on $S$. Then,
there exists physical (normalizable) $\cN$-fold SUSY states $|\Psi_{0}^{-}
\rangle$ and/or $|\Psi_{0}^{+}\rangle$ which satisfies (\ref{eq:NSsts}) if
$\cV_{\cN}^{-}(S)\subset L^{2}(S)$ and/or $\cV_{\cN}^{+}(S)\subset L^{2}(S)$,
in other words, if $H^{-}$ and/or $H^{+}$ is \emph{quasi-exactly solvable}.
If there are no such physical $\cN$-fold SUSY states in the Hilbert space
$L^2(S)$ exists then $\cN$-fold SUSY of the system is said to be dynamically
broken. It was first shown correctly in Ref.~\cite{AST01b} that the generalized
Witten index characterizes $\cN$-fold SUSY breaking, which corrected the wrong
statement made earlier in Ref.~\cite{AIS93}.

For $\cN>1$, we can have an intriguing situation where not the whole
of, but a subspace of the solvable sectors $\cV_{\cN}^{-}(S)$ and/or
$\cV_{\cN}^{+}(S)$ belong to the Hilbert space $L^{2}(S)$. In this case,
$\cN$-fold SUSY of the system is said to be \emph{partially broken}.
Partial breaking of $\cN$-fold SUSY was first discovered in Ref.~\cite{GT05}.
We note that it is different in nature from the partial breaking of (nonlinear)
SUSY~\cite{BW84,Ba85}.

Construction of an $\cN$-fold SUSY system is in general quite difficult,
especially for a larger value of $\cN$, since the intertwining relations
(\ref{eq:inter}) compose of coupled nonlinear differential equations for
$U^{\pm}(q)$ and $w_{k}^{[\cN]}(q)$ ($k=0,\dots,\cN-1$). For the direct
calculations of intertwining relations in a PDM background in the cases
of $\cN=1$ and $2$, see Ref.~\cite{MRR10}. To circumvent the difficulty,
a systematic algorithm for constructing an $\cN$-fold SUSY system was
developed in Ref.~\cite{GT05} for constant-mass quantum mechanics and
was later generalized to PDM systems in Ref.~\cite{Ta06a}. The significant
feature which is common in both constant-mass and PDM systems is that
an $\cN$-dimensional linear space of functions
\begin{align}
\tcV_{\cN}^{-}=\braket{\tvph_{1}(z),\dots,\tvph_{\cN}(z)},
\end{align}
preserved by a second-order linear differential operator $\tH^{-}$
can determine whole of an $\cN$-fold SUSY system. Indeed, we can construct
a pair of $\cN$th-order linear differential operators $\bar{\tP}_{\cN}^{\pm}$
and another $\cN$-dimensional vector space $\bar{\cV}_{\cN}^{+}$ such that
$\bar{\tcV}_{\cN}^{\pm}=\ker\bar{\tP}_{\cN}^{\pm}$. Then, we can show that
a pair of second-order linear differential operators given by
\begin{align}
\bar{\tH}^{\pm}=&-A(z)\frac{\rmd^{2}}{\rmd z^{2}}+\left[\frac{\cN-2}{2}
 A'(z)\pm Q(z)\right]\frac{\rmd}{\rmd z}-C(z)\notag\\
&-(1\pm1)\left[\frac{\cN-1}{2}Q'(z)-\frac{1}{2}A'(z)\tw_{\cN-1}^{[\cN]}
 (z)-A(z)\tw_{\cN-1}^{[\cN]\prime}(z)\right],
\label{eq:btH+-}
\end{align}
is weakly quasi-solvable with respect to the spaces $\bar{\tcV}_{\cN}^{\pm}$,
namely, $\bar{\tH}^{\pm}\bar{\tcV}_{\cN}^{\pm}\subset\bar{\tcV}_{\cN}^{\pm}$.

With the choice of the change of variable $z=z(q)$ and the gauge potential
$\cW_{\cN}^{\pm}$ determined by
\begin{align}
z'(q)^{2}=2m(q)A(z),\quad\cW_{\cN}^{\pm}=-\frac{1}{4}\ln|m(q)|+\frac{\cN-1}{4}
 \ln|2A(z)|\pm\int\rmd z\,\frac{m(q)Q(z)}{2A(z)},
\label{eq:gauge}
\end{align}
we can obtain an $\cN$-fold SUSY system by
\begin{align}
H^{\pm}=\rme^{-\cW_{\cN}^{\pm}}\bar{\tH}^{\pm}\rme^{\cW_{\cN}^{\pm}}
 \Bigr|_{z=z(q)},\qquad P_{\cN}^{\pm}=\rme^{-\cW_{\cN}^{\pm}}
 \bar{\tP}_{\cN}^{\pm}\rme^{\cW_{\cN}^{\pm}}\Bigr|_{z=z(q)}.
\label{eq:gtf2}
\end{align}
With the change of variable and the gauge transformation, both of $H^{\pm}$
get the form of PDM Hamiltonian (\ref{eq:H+-}) and their effective potentials
$U^{\pm}(q)$ are given by
\begin{align}
U^{\pm}(q)=&\;\frac{1}{2m(q)}\left[\left(\frac{\rmd\cW_{\cN}^{-}}{\rmd q}
 \right)^{2}-\frac{\rmd^{2}\cW_{\cN}^{-}}{\rmd q^{2}}+\frac{m'(q)}{m(q)}
 \frac{\rmd\cW_{\cN}^{-}}{\rmd q}\right]-C(z(q))\notag\\
&-(1\pm1)\left[\frac{\cN-1}{2}Q'(z)-\frac{1}{2}A'(z)\tw_{\cN-1}^{[\cN]}(z)
 -A(z)\tw_{\cN-1}^{[\cN]\prime}(z)\right]_{z=z(q)}.
\label{eq:U+-}
\end{align}
The solvable sectors $\cV_{\cN}^{\pm}$ of $H^{\pm}$ are evidently given by
\begin{align}
\cV_{\cN}^{\pm}=\ker P_{\cN}^{\pm}=\rme^{-\cW_{\cN}^{\pm}}
 \bar{\tcV}_{\cN}^{\pm}\Bigr|_{z=z(q)}.
\label{eq:VN+-}
\end{align}
In principle, we can construct a pair of $\cN$-fold SUSY PDM Hamiltonians
$H^{\pm}$ and its solvable sectors $\cV_{\cN}^{\pm}$ by using the formulas
(\ref{eq:U+-}) and (\ref{eq:VN+-}). However, there
is an easier way to obtain such a system when we have already had
an ordinary $\cN$-fold SUSY constant-mass quantum system at hand. Suppose
the latter system is such that its pair of potentials $V^{(0)\pm}(q)$,
its gauge potentials $\cW_{\cN}^{(0)\pm}(q)$, its solvable sectors
$\cV_{\cN}^{(0)\pm}[q]$ are all known. Then, an $\cN$-fold SUSY PDM
system having a pair of effective potentials $U^{\pm}(q)$, gauge
potentials $\cW_{\cN}^{\pm}(q)$, and solvable sectors $\cV_{\cN}^{\pm}
[q]$ can be constructed immediately via the following prescription:
\begin{subequations}
\begin{align}
U^{\pm}(q)&=V^{(0)\pm}(u(q))+\frac{m''(q)}{8m(q)^{2}}-\frac{7m'(q)^{2}}{
 32m(q)^{3}},
\label{eq:PCT1}\\
\cW_{\cN}^{\pm}(q)&=-\frac{1}{4}\ln|m(q)|+\cW_{\cN}^{(0)\pm}(u(q)),\\[5pt]
\cV_{\cN}^{\pm}[q]&=m(q)^{1/4}\cV_{\cN}^{(0)\pm}[u(q)],
\label{eq:PCT3}
\end{align}\label{e44}
\end{subequations}
where the function $u(q)$ is given by
\begin{align}
u(q)=\int\rmd q\sqrt{m(q)}.
\label{eq:uq}
\end{align}
Actually, the above relations are consistent with the formulas obtained
by the point canonical transformation, see, e.g., equations (2.7) and
(2.8) in Ref.~\cite{Al02}, equation (7) of \cite{RR01} and equations (10),
(13), and (14) in Ref.~\cite{GOGU02}. The above relations (\ref{e44}) have
also been verified in Ref.~\cite{Ta06a} where type A $\cN$-fold SUSY has
been constructed in PDM background.
One of the most salient features unveiled by the algorithmic construction
is that both constant-mass and PDM quantum systems with $\cN$-fold SUSY
have totally the same structure in the gauged $z$-space. That is, the
functional forms of the gauged operators such as $\bar{\tP}_{\cN}^{\pm}$ and
$\bar{\tH}^{\pm}$ given by (\ref{eq:btH+-})
are identical in both the cases. It means in
particular that the starting vector space $\tcV_{\cN}^{-}$ determines
all in the algorithm regardless of whether mass is constant or not.
Hence, different types of $\cN$-fold SUSY are characterized by different
types of vector spaces $\tcV_{\cN}^{-}$ and vice versa. Until now, four
different types have been discovered, namely, type A~\cite{AST01a,Ta03a},
type B~\cite{GT04}, type C~\cite{GT05}, and type $X_{2}$~\cite{Ta10a}.
We note that almost all the models having essentially the same symmetry as
$\cN$-fold SUSY but called with other terminologies in the literature, such as
P\"{o}schl--Teller and Lam\'{e} potentials, are actually particular cases of type
A $\cN$-fold SUSY.
In this article, we focus on constructing PDM quantum systems with
type B and type $X_{2}$ $\cN$-fold SUSY since the other types (type A and
type C) are not related to exceptional polynomial subspaces. In what follows, we
shall review the general structure of these two types of $\cN$-fold SUSY.

\subsection{Type B $\cN$-fold Supersymmetry}

Type B $\cN$-fold SUSY was first discovered in Ref.~\cite{GT04} and
was found to be associated with the following monomial space
\begin{align}
\tcV_{\cN}^{-}=\tcV_{\cN}^{(\mathrm{B})}:=\braket{1,z,\dots,z^{\cN-2},
 z^{\cN}},
\label{eq:tVB-}
\end{align}
called type B, which was considered in Ref.~\cite{PT95} in the context of
the classification of monomial spaces preserved by second-order linear
ordinary differential operators. Applying the algorithm to the type B
monomial space, we obtain~\cite{Ta09} the gauged $\cN$-fold supercharge
components
\begin{align}
\tP_{\cN}^{-}=z'(q)^{\cN}\left(\frac{\rmd}{\rmd z}-\frac{1}{z}\right)
 \frac{\rmd^{\cN-1}}{\rmd z^{\cN-1}},\qquad\bar{P}_{\cN}^{+}=z'(q)^{\cN}
 \frac{\rmd^{\cN-1}}{\rmd z^{\cN-1}}\left(\frac{\rmd}{\rmd z}+\frac{1}{z}
 \right),
\end{align}
and the functions which characterize the gauged Hamiltonians (\ref{eq:btH+-})
are given by
\begin{align}
A(z)&=a_{4}z^{4}+a_{3}z^{3}+a_{2}z^{2}+a_{1}z+a_{0},\\
2Q(z)&=-\cN a_{3}z^{2}+2b_{1}z-\cN a_{1},\\
C(z)&=\cN(\cN-3)a_{4}z^{2}+\cN(\cN-2)a_{3}z+c_{0},
\end{align}
and $\tw_{\cN-1}^{[\cN]}(z)=-z^{-1}$. The other linear space $\bar{\cV}_{\cN}^{+}$
preserved by $\bar{H}^{+}$ is given by
\begin{align}
\bar{\cV}_{\cN}^{+}=z^{-1}\braket{1,z^{2},\dots,z^{\cN}}.
\label{eq:bVB+}
\end{align}
We note that both the monomial spaces (\ref{eq:tVB-}) and (\ref{eq:bVB+})
are actually exceptional polynomial subspaces of codimension $1$, see
Ref.~\cite{GKM08a}.
We can easily check that the type B Hamiltonian $H^{+}$ preserves an infinite
flag of the following spaces
\begin{align}
\bar{\cV}_{1}^{+}\rme^{-\cW_{\cN}^{+}}\subset\bar{\cV}_{2}^{+}
 \rme^{-\cW_{\cN}^{+}}\subset\cdots\subset\bar{\cV}_{\cN}^{+}
 \rme^{-\cW_{\cN}^{+}}\subset\cdots,
\end{align}
where $\bar{\cV}_{\cN}^{+}$ and $\cW_{\cN}^{+}$ are given by (\ref{eq:bVB+})
and (\ref{eq:gauge}), respectively, and thus $H^{+}$ is solvable if and only if
$a_{3}=a_{4}=0$. On the other hand, the partner type B Hamiltonian $H^{-}$
does not appear to be solvable for any parameter value since the type B
monomial space (\ref{eq:tVB-}) does not constitutes an infinite flag due to
the fact that $\tcV_{\cN}^{(\mathrm{B})}\not\subset\tcV_{\cN+1}^{(\mathrm{B})}$
for all $\cN=1,2,\ldots$. However, it turns out \cite{Ta09} that, when
$a_{3}=a_{4}=0$ and $H^{+}$ gets solvable, the partner Hamiltonian $H^{-}$
does preserve an infinite flag of linear spaces given by
\begin{align}
\tcV_{1}^{(\mathrm{A})}\rme^{-\cW_{\cN}^{-}}\subset\tcV_{2}^{(\mathrm{A})}
 \rme^{-\cW_{\cN}^{-}}\subset\cdots\subset\tcV_{\cN}^{(\mathrm{A})}
 \rme^{-\cW_{\cN}^{-}}\subset\cdots,
\end{align}
where $\cW_{\cN}^{-}$ is given by (\ref{eq:gauge}) and $\tcV_{\cN}^{(\mathrm{A})
}$ is the type A monomial space defined by
\begin{align}
\tcV_{\cN}^{(\mathrm{A})}=\braket{1,z,\dots,z^{\cN-1}}.
\end{align}
That is, $H^{-}$ and $H^{+}$ can be solvable simultaneously. In this paper,
all the type B models we will consider later satisfy the solvability
condition $a_{3}=a_{4}=0$. Thus, all the pairs of type B Hamiltonians
$H^{\pm}$ preserve the infinite-dimensional solvable sectors
$\cV^{\pm}$ given by
\begin{align}
\begin{split}
\cV^{-}&=\braket{1,z(q),z(q)^{2},\ldots}\,\rme^{-\cW_{\cN}^{-}(q)},\\
\cV^{+}&=\braket{1,z(q)^{2},z(q)^{3},\ldots}z(q)^{-1}\rme^{
 -\cW_{\cN}^{+}(q)}.
\label{eq:Bidss}
\end{split}
\end{align}
An interesting consequence of the fact that $H^{-}$ and $H^{+}$ preserve
different types of infinite flag of spaces in the solvable case is that
the eigenfunctions of $H^{-}$ are expressed in terms of a classical
polynomial system while those of $H^{+}$ are in terms of an $X_{1}$
polynomial system. It is exactly the underlying reason why some of the
Hamiltonians whose eigenfunctions are expressed in terms of the $X_{1}$
Laguerre or Jacobi polynomials were obtained by those whose bound state
eigenfunctions are expressed in terms of the classical Laguerre or
Jacobi polynomials using an intertwining or SUSY techniques in
Refs.~\cite{BQR09,CKS95,BGQR05}.

\subsection{Type $X_{2}$ $\cN$-fold Supersymmetry}

Type $X_{2}$ $\cN$-fold SUSY constructed in Ref.~\cite{Ta10a} is associated
with the following exceptional polynomial subspace of codimension $2$
\begin{align}
\tcV_{\cN}^{-}=\braket{\tvph_{1}(z;\alpha),\dots,\tvph_{\cN}(z;\alpha)},
\label{eq:tVX-}
\end{align}
where $\tvph_{n}(z;\alpha)$ is a polynomial of degree $n+1$ in $z$ with
a parameter $\alpha(\neq0,1)$ defined by
\begin{align}
\tvph_{n}(z;\alpha)=(\alpha+n-2)z^{n+1}+2(\alpha+n-1)(\alpha-1)z^{n}
 +(\alpha+n)(\alpha-1)\alpha z^{n-1}.
\end{align}
Applying the algorithm to the $X_{2}$ space (\ref{eq:tVX-}), we
obtain~\cite{Ta10a} the gauged $\cN$-fold supercharge components
\begin{align}
\begin{split}
\tP_{\cN}^{-}&=z'(q)^{\cN}\frac{f(z;\alpha)}{f(z;\alpha+\cN)}
 \prod_{k=0}^{\cN-1}\frac{f(z;\alpha+k+1)}{f(z;\alpha+k)}\left(
 \frac{\rmd}{\rmd z}-\frac{f'(z;\alpha+k+1)}{f(z;\alpha+k+1)}\right),\\
\bar{P}_{\cN}^{+}&=z'(q)^{\cN}\left[\prod_{k=0}^{\cN-1}\left(
 \frac{\rmd}{\rmd z}+\frac{f'(z;\alpha+\cN-k)}{f(z;\alpha+\cN-k)}\right)
 \frac{f(z;\alpha+\cN-k)}{f(z;\alpha+\cN-k-1)}\right]
 \frac{f(z;\alpha)}{f(z;\alpha+\cN)},
\end{split}
\end{align}
where $\prod_{k=0}^{\cN-1}A_{k}:=A_{\cN-1}\dots A_{1}A_{0}$, and the functions
$f(z;\alpha)$ and $\tw_{\cN-1}^{[\cN]}(z)$ are given by
\begin{align}
f(z;\alpha)&=z^{2}+2(\alpha-1)z+(\alpha-1)\alpha,\\
\tw_{\cN-1}^{[\cN]}(z)&=-(\cN-1)\frac{f'(z;\alpha)}{f(z;\alpha)}
 -\frac{f'(z;\alpha+\cN)}{f(z;\alpha+\cN)}.
\label{eq:X2tw}
\end{align}
The most general forms of the functions $A(z)$, $Q(z)$, and $C(z)$ appeared
in $\bar{\tH}^{\pm}$ depend
on four parameters $a_{i}$ ($i=1,\dots,4$), but in this paper we only
consider models with $a_{4}=a_{3}=0$. In the latter case, they
read as
\begin{align}
A(z)=&\;a_{2}z^{2}+a_{1}z+(\alpha-1)(\alpha+\cN-1)a_{2},\\[5pt]
Q(z)=&-a_{2}z^{2}-(3a_{2}+a_{1})z-(\alpha-1)(3\alpha+3\cN-7)a_{2}\notag\\
&+\frac{2\alpha+\cN-8}{2}a_{1}+\frac{4(\alpha-1)D(z)}{f(z;\alpha)},\\
C(z)=&\;a_{2}z+c_{0}-\frac{4(\alpha-1)D(z)}{f(z;\alpha)},
\label{eq:defC}
\end{align}
where $D(z)$ is given by
\begin{align}
D(z)=-[(2\alpha+\cN-3)a_{2}-a_{1}]z-(\alpha-1)(2\alpha+\cN-1)a_{2}
 +\alpha a_{1}.
\end{align}
For their most general forms, please refer to Ref.~\cite{Ta10a}.
The other linear space $\bar{\cV}_{\cN}^{+}$ preserved by $\bar{H}^{+}$
is given by
\begin{align}
\bar{\cV}_{\cN}^{+}=\braket{\bar{\chi}_{1}(z;\alpha+\cN),\dots,
 \bar{\chi}_{\cN}(z;\alpha+\cN)}f(z;\alpha)^{-1}f(z;\alpha+\cN)^{-1},
\end{align}
where $\bar{\chi}_{n}(z;\alpha)$ is a polynomial of degree $n+1$ in $z$
defined by
\begin{align}
\bar{\chi}_{n}(z;\alpha)=&\;(\alpha-n)(\alpha-n+1)z^{n+1}+2(\alpha-n-1)
 (\alpha-n+1)(\alpha-1)z^{n}\notag\\
&+(\alpha-n-1)(\alpha-n)(\alpha-1)\alpha z^{n-1}.
\end{align}
The solvable sectors
$\cV_{\cN}^{\pm}$ of the constant-mass Hamiltonians $H^{\pm}$ are
\begin{align}
\cV_{\cN}^{-}&=\braket{\tvph_{1}(z(q);\alpha),\dots,\tvph_{\cN}(z(q);
 \alpha)}\,\rme^{-\cW_{\cN}^{-}(q)},\\
\cV_{\cN}^{+}&=\frac{\braket{\bar{\chi}_{1}(z(q);\alpha+\cN),\dots,
 \bar{\chi}_{\cN}(z(q);\alpha+\cN)}}{f(z(q);\alpha)f(z(q);\alpha+\cN)}
 \,\rme^{-\cW_{\cN}^{+}(q)}.
\end{align}
Finally, the type $X_{2}$ Hamiltonians $H^{\pm}$ preserve the infinite
flag of the spaces $\cV_{\cN}^{\pm}$ ($\cN=1,2,\ldots$) and are
simultaneously solvable if and only if $a_{2}=(a_{3}=a_{4}=)0$.

\section{Type B and Type $X_2$ $\cN$-fold Supersymmetry for position-dependent
 mass}\label{s3}

In this section, we shall consider some models which belong to type B and
type $X_{2}$ $\cN$-fold supersymmetry. In order to study effect of PDM
in these models, we need to consider simultaneously the corresponding
constant-mass type B and type $X_{2}$ models as well. In particular, we shall
address ourselves to the following question: Does position dependent mass
have any effect on dynamical breaking of type B and type $X_{2}$ $\cN$-fold SUSY?
By comparing the solvable sectors of both constant and position-dependent
mass cases, we shall see below that the answer is in the affirmative in some
cases for particular choices of physically interesting mass functions.
In order to explore in detail the impact of mass functions on symmetry
breaking or restoration, it will be appropriate to consider more than one
mass function in a few examples. Also it will be shown that
the bound state wavefunctions of one of the partner potentials obtained in
type B $\cN$-fold SUSY are associated with exceptional $X_{1}$
Laguerre and Jacobi polynomials while those of the other partner are
associated with classical Laguerre and Jacobi polynomials.

\subsection{Effects of PDM on Dynamical Symmetry Breaking of Type B
 $\cN$-fold SUSY}\label{ss3.1}

Here we shall consider three examples of type B $\cN$-fold SUSY corresponding
to three different choices of $A(z)$. In each of the examples, we first
show the results in the constant mass case, followed by the corresponding
results in the PDM case. As we referred to before, all the type B models
constructed below satisfy the solvability condition $a_{3}=a_{4}=0$ and
thus their solvable sectors in the constant-mass case are given by
(\ref{eq:Bidss}).\\

\noindent
\textbf{Example 3.1.} $A(z)=k(z-z_{0})\quad(k\neq0)$\\
Potentials:
\begin{align}
V^{(0)-}(q)&=\frac{b_{1}^{\,2}}{8}q^{2}+\frac{4(z_{0}b_{1}-\cN k)^{2}
 -k^{2}}{8 k^{2}q^{2}}+\frac{\cN b_{1}}{2}+V_{0},
\label{e11}\\
V^{(0)+}(q)&=\frac{b_{1}^{\,2}}{8}q^{2}+\frac{4z_{0}^{\,2}b_{1}^{\,2}
 -k^{2}}{8 k^{2}q^{2}}+\frac{2k}{k q^{2}+2 z_{0}}
 -\frac{8k z_{0}}{(k q^{2}+2 z_{0})^{2}}+V_{0},
\label{e21}
\end{align}
where $V_{0}$ is an irrelevant constant given by
\begin{align*}
V_{0}=\frac{(z_{0}b_{1}-\cN k)b_{1}}{2k}+\frac{b_{1}}{\cN}-R.
\end{align*}
Solvable sectors:
\begin{align}
\cV^{(0)-}&=\braket{1,z(q),z(q)^{2},\ldots}
 q^{(2z_{0}b_{1}-2\cN k+k)/(2k)}\rme^{b_{1}q^{2}/4},
\label{e12}\\
\cV^{(0)+}&=\braket{1,z(q)^{2},z(q)^{3},\ldots}z(q)^{-1}
 q^{-(2z_{0}b_{1}-k)/(2k)}\rme^{-b_{1}q^{2}/4}.
\label{e13}
\end{align}
We assume $k>0$ and $z_{0}>0$ so that the pair of potentials $V^{\pm}
(q)$ has no singularities except for at $q=0$. Thus, the system is naturally
defined in $L^{2}(\bbR_{+})$, $\bbR_{+}=(0,\infty)$. In the latter Hilbert
space, $\cV^{(0)-}(\bbR_{+})\subset L^{2}(\bbR_{+})$ if and only if
\begin{align}
b_{1}<0\quad\text{and}\quad z_{0}b_{1}>(\cN-1)k,
\end{align}
which cannot be satisfied by any $b_{1}\in\bbR$. On the other hand,
$\cV^{(0)+}(\bbR_{+})\subset L^{2}(\bbR_{+})$ if and only if
\begin{align}
b_{1}>0\quad\text{and}\quad k>z_{0}b_{1}.
\end{align}
Hence, $\cN$-fold SUSY of the system is unbroken if and only if
$0<b_{1}<k/z_{0}$ on the constant-mass background.

Now, the relevant expressions for partner potentials, gauge potentials and
corresponding solvable sectors of type B PDM systems can be obtained using
Eqs.~(\ref{e11})--(\ref{e13}) and relations (\ref{e44}). Since
our main objective in this section is to study effect of mass function
on dynamical breaking of $\cN$-fold SUSY, we give below only the solvable
sectors $\cV^{\pm}$ for an arbitrary mass function $m(q)$:
\begin{align}
\cV^{-}&=\braket{1,z(u(q)),z(u(q))^{2},\dots}
 m(q)^{1/4}u(q)^{(2z_{0}b_{1}-2\cN k+k)/(2k)}\rme^{b_{1}u(q)^{2}/4},
\label{e60}\\
\cV^{+}&=\braket{1,z(u(q))^{2},z(u(q))^{3},\dots}z(u(q))^{-1}
 m(q)^{1/4}u(q)^{-(2z_{0}b_{1}-k)/(2k)}\rme^{-b_{1}u(q)^{2}/4},
\label{e61}
\end{align}
where $u(q)$ is given by (\ref{eq:uq}). At this point, we are in a position
to choose a particular mass function. Let the mass function be
\begin{align}
m(q)=\rme^{-bq},\quad b>0,\quad q\in(-\infty,\infty),
\end{align}
which was considered in Ref.~\cite{MR09a} where the PDM potentials were
associated with $X_{1}$-Laguerre polynomials. This exponentially behaved mass
function has been often used in the study of confined energy states for carriers
in semiconductor quantum well~\cite{GOGU02,MR09a}. It has also been used to
compute transmission probabilities for scattering in abrupt heterostructures
\cite{KKS05} which may be useful in the design of semiconductor devices
\cite{CHA90}. For the mass function, the change of variable is given by
\begin{align}
u(q)=-\frac{2}{b}\rme^{-bq/2},
\end{align}
and the pair of potentials $U^{\pm}(q)$ reads from (\ref{eq:PCT1}) as
\begin{align}
U^{-}(q)&=\frac{b_{1}^{\,2}}{2b^{2}}\rme^{-bq}+\frac{b^{2}[(z_{0}b_{1}
 -\cN k)^{2}-k^{2}]}{8k^{2}}\rme^{bq}+\frac{\cN b_{1}}{2}+V_{0},
\label{e22}\\
U^{+}(q)&=\frac{b_{1}^{\,2}}{2b^{2}}\rme^{-bq}+\frac{b^{2}(z_{0}^{\,2}
 b_{1}^{\,2}-k^{2})}{8k^{2}}\rme^{bq}+\frac{k b^{2}}{2k\rme^{-bq}
 +z_{0}b^{2}}-\frac{2kz_{0}b^{4}}{(2k\rme^{-bq}+z_{0}b^{2})^{2}}+V_{0},
\label{e24}
\end{align}
respectively. It is worth mentioning here that the potential $U^{+}(q)$
given in (\ref{e22}) is identical with the potential $V_{eff}(q)$ associated
with exceptional $X_{1}$ Laguerre polynomials [e.g., Eq.~(12) of
Ref.~\cite{MR09a}], if one takes  $k=1/2$, $b_{1}=b^{2}/2$, and $z_{0}=
\alpha/b^{2}$. On the other hand, for the same choices of parameters
the other potential $U^{-}(q)$ coincides with the potential [after making a
translation $\alpha\to\alpha-\cN$] previously obtained in Ref.~\cite{BGQR05}
corresponding to classical Laguerre polynomials.

The solvable sectors of the potentials (\ref{e22}) and (\ref{e24}) are
respectively given by
\begin{align}
\cV^{-}=&\;\braket{1,\rme^{-bq}+\bar{z}_{0},
 (\rme^{-bq}+\bar{z}_{0})^{2},\ldots}\notag\\
&\times\exp\left[-\left(\frac{z_{0}b_{1}}{k}-\cN+1\right)\frac{b}{2}q
 +\frac{b_{1}}{b^{2}}\rme^{-bq}\right],
\label{e25}
\end{align}
\begin{align}
\cV^{+}=&\;\braket{1,(\rme^{-bq}+\bar{z}_{0})^{2},
 (\rme^{-bq}+\bar{z}_{0})^{3},\ldots}\notag\\
&\times(\rme^{-bq}+\bar{z}_{0})^{-1}\exp\left[\left(\frac{z_{0}b_{1}}{k}-1
 \right)\frac{b}{2}q-\frac{b_{1}}{b^{2}}\rme^{-bq}\right],
\label{e26}
\end{align}
where $\bar{z}_{0}=z_{0}b^{2}/(2k)$. Here the potentials have no
singularities in the finite part of the real line, so the domain is
$\bbR$. Since $b>0$, so $\cV^{-}(\bbR)\subset L^{2}(\bbR)$ if and only
if $b_{1}<0$. On the other hand, $\cV^{+}(\bbR)\subset L^{2}(\bbR)$ if
and only if $b_{1}>0$. Hence, the $\cN$-fold SUSY of the PDM system is
unbroken unless $b_{1}=0$.
Comparing the solvable sectors of both the constant and position-dependent
mass scenarios, it can be observed that it is not possible to break
$\cN$-fold SUSY dynamically for the particular choice of mass function
$m(q)=\rme^{-bq}$. In addition, we have checked that many physically
interesting mass functions also have no effect on symmetry breaking.\\

\noindent
\textbf{Example 3.2.} $A(z)=a^{2}[1-(z-z_{0})^{2}]/2\quad(a>0)$\\
Potentials:
\begin{align}
V^{(0)-}(q)=&\;\frac{(4b_{1}^{\,2}-\cN^{\,2}a^{4})z_{0}}{4a^{2}}
 \frac{\sin aq}{\cos^{2}aq}+\frac{(2b_{1}-\cN a^{2})^{2}z_{0}^{\,2}
 +(2b_{1}+\cN a^{2})^{2}-a^{4}}{8 a^{2}}\tan^{2}aq\notag\\
&+\frac{b_{1}\cN}{2}+V_{0},
\label{e14}
\end{align}
\begin{align}
V^{(0)+}(q)=&\;\frac{(2b_{1}-\cN a^{2})^{2}z_{0}}{4 a^{2}}
 \frac{\sin aq}{\cos^{2}aq}+\frac{(2b_{1}-\cN a^{2})^{2}(z_{0}^{\,2}
 +1)-a^{4}}{8a^{2}}\tan^{2}aq\notag\\
&+\frac{a^{2}z_{0}}{\sin aq+z_{0}}-\frac{a^{2}(z_{0}^{\,2}-1)}{
 (\sin aq+z_{0})^{2}}-\frac{b_{1}\cN}{2}+V_{0},
\label{e15}
\end{align}
where $V_{0}$ is an irrelevant constant given by
\begin{align*}
V_{0}=\frac{b_{1}}{\cN}+\frac{a^{2}(\cN^{2}-7)}{12}
 +\frac{(2b_{1}z_{0}-\cN z_{0}a^{2})^{2}}{8a^{2}}-R.
\end{align*}
Solvable sectors:
\begin{align}
\cV^{(0)-}&=\braket{1,z(q),z(q)^{2},\ldots}|\cos aq|^{\frac{b_{1}}{a^{2}}
 +\frac{\cN-1}{2}}\left(\frac{1+\sin aq}{1-\sin aq}
 \right)^{-\frac{(2b_{1}-\cN a^{2})z_{0}}{4a^{2}}},
\label{e16}\\
\cV^{(0)+}&=\braket{1,z(q)^{2},z(q)^{3},\ldots}z(q)^{-1}
 |\cos aq|^{-\frac{b_{1}}{a^{2}}+\frac{\cN-1}{2}}\left(
 \frac{1+\sin aq}{1-\sin aq}\right)^{\frac{(2b_{1}-\cN a^{2})z_{0}}{4a^{2}}}.
\label{e17}
\end{align}
It is worth mentioning here that the potential $V^{(0)+}(q)$ coincides with
the potential whose bound state wave functions are given in terms of
exceptional $X_{1}$ Jacobi polynomial \cite{Qu08b} for $a=1$, $b_{1}=B+\cN/2$,
$z_{0}=-(2A-1)/(2B)$ whereas potential $V^{(0)-}(q)$ coincides with the
Scarf~I potential \cite{CKS95} [after making an change $B\to B+\cN$] whose
bound state wave functions are given in terms of classical Jacobi polynomials.

We choose here a domain of the system as $S=(-\frac{\pi}{2a},\frac{\pi}{2a})$
and assume $z_{0}>1$ so that the pair of potentials $V^{(0)\pm}(q)$ has no
singularities except for at the boundary $\partial S=\{-\frac{\pi}{2a},
\frac{\pi}{2a}\}$. Thus, the Hilbert space for the system is $L^{2}(S)$.
Then, $\cV^{(0)-}(S)\subset L^{2}(S)$ if and only if
\begin{align*}
\frac{b_{1}}{a^{2}}+\frac{\cN-1}{2}\pm\frac{(2b_{1}-\cN a^{2})z_{0}}{2a^{2}}
 >-\frac{1}{2},
\end{align*}
that is,
\begin{align}
\frac{\cN a^{2}}{2}\frac{z_{0}-1}{z_{0}+1}<b_{1}<\frac{\cN a^{2}}{2}
 \frac{z_{0}+1}{z_{0}-1}\quad\text{for}\quad z_{0}>1.
\end{align}
Similarly, $\cV^{(0)+}(S)\subset L^{2}(S)$ if and only if
\begin{align*}
-\frac{b_{1}}{a^{2}}+\frac{\cN-1}{2}\pm\frac{(2b_{1}-\cN a^{2})z_{0}}{2
 a^{2}}>-\frac{1}{2},
\end{align*}
that is,
\begin{align}
b_{1}>\frac{\cN a^{2}}{2}\quad\text{and}\quad z_{0}>1.
\end{align}
Hence, $\cN$-fold SUSY of the system is broken for the constant mass case
if and only if $z_{0}>1$ and
\begin{align}
b_{1}\leq\frac{\cN a^{2}}{2}\frac{z_{0}-1}{z_{0}+1}\quad\text{or}\quad
 b_{1}\geq\frac{\cN a^{2}}{2}\frac{z_{0}+1}{z_{0}-1}.
\end{align}
In a PDM case, the solvable sectors $\cV^{\pm}$ of the type B PDM
$\cN$-fold SUSY partner Hamiltonians $H^{\pm}$ for an arbitrary mass
function $m(q)$ are deformed according to (\ref{eq:PCT3}) as
\begin{align}
\cV^{-}=&\;\braket{1,z(u(q)),z(u(q))^{2},\ldots}m(q)^{\frac{1}{4}}\notag\\
&\times|\cos a u(q)|^{\frac{b_{1}}{a^{2}}+\frac{\cN-1}{2}}\left(
 \frac{1+\sin a u(q)}{1-\sin a u(q)}\right)^{-\frac{(2b_{1}-\cN
 a^{2})z_{0}}{4a^{2}}},
\label{e62}
\end{align}
\begin{align}
\cV^{+}=&\;\braket{1,z(u(q))^{2},z(u(q))^{3},\ldots}
 m(q)^{\frac{1}{4}}\notag\\
&\times\frac{|\cos a u(q)|^{-\frac{b_{1}}{a^{2}}+\frac{\cN-1}{2}}}{\sin a
 u(q)+z_{0}}\left(\frac{1+\sin a u(q)}{1-\sin a u(q)}\right)^{\frac{(2b_{1}
 -\cN a^{2})z_{0}}{4a^{2}}}.
\label{e63}
\end{align}
where $u(q)$ is given by (\ref{eq:uq}). In this case, the choice of mass
function and the corresponding change of variable are given by
\begin{align}
m(q)=\frac{2}{\pi}\rme^{-2q^{2}},\qquad u(q)=\ef q,
 \qquad q\in(-\infty,\infty).
\label{e66}
\end{align}
Consequently, the partner potentials $U^{\pm}(q)$ read as
\begin{align}
U^{-}(q)=&\;\frac{(4b_{1}^{\,2}-\cN^{\,2}a^{4})z_{0}}{4a^{2}}\frac{\sin
 (a\ef q)}{\cos^{2}(a\ef q)}-\frac{(3q^{2}+1)\pi\rme^{2q^{2}}}{4}
 +\frac{b_{1}\cN}{2}+V_{0}\notag\\
&+\frac{(2b_{1}-\cN a^{2})^{2}z_{0}^{\,2}+(2b_{1}+\cN a^{2})^{2}-a^{4}}{8
 a^{2}}\tan^{2}(a\ef q),
\label{e64}
\end{align}
\begin{align}
U^{+}(q)=&\;\frac{(2b_{1}-\cN a^{2})^{2}z_{0}}{4a^{2}}\frac{\sin
 (a\ef q)}{\cos^{2}(a\ef q)}-\frac{(3q^{2}+1)\pi\rme^{2q^{2}}}{4}
 -\frac{b_{1}\cN}{2}+V_{0}\notag\\
&+\frac{a^{2}z_{0}}{\sin(a\ef q)+z_{0}}-\frac{a^{2}(z_{0}^{\,2}-1)}{
 [\sin(a\ef q)+z_{0}]^{2}}\notag\\
&+\frac{(2b_{1}-\cN a^{2})^{2}(z_{0}^{\,2}+1)-a^{4}}{8a^{2}}\tan^{2}(a\ef q).
\label{e65}
\end{align}
The solvable sectors of the potentials (\ref{e64}) and (\ref{e65}) are
given by
\begin{align}
\cV^{-}=&\;\braket{1,z(u(q)),z(u(q))^{2},\ldots}\,\rme^{-q^{2}/4}\notag\\
&\times|\cos(a\ef q)|^{\frac{b_{1}}{a^{2}}+\frac{\cN-1}{2}}
 \left(\frac{1+\sin(a\ef q)}{1-\sin(a\ef q)}\right)^{
 -\frac{(2b_{1}-\cN a^{2})z_{0}}{4a^{2}}},
\label{e67}
\end{align}
\begin{align}
\cV^{+}=&\;\braket{1,z(u(q))^{2},z(u(q))^{3},\ldots}\,\rme^{-q^{2}/4}\notag\\
&\times\frac{|\cos(a\ef q)|^{-\frac{b_{1}}{a^{2}}+\frac{\cN-1}{2}}}{
 \sin(a\ef q)+z_{0}}\left(\frac{1+\sin(a\ef q)}{1-\sin(a\ef q)}
 \right)^{\frac{(2b_{1}-\cN a^{2})z_{0}}{4a^{2}}}.
\label{e68}
\end{align}
The potentials $U^{\pm}(q)$ as well as the mass function are well behaved
in $q\in(-\infty,\infty)$. So, we can take the domain as the whole real
line $\bbR$. Since $\ef q\to\pm1$ as $q\to\pm\infty$, so both the solvable
sectors $\cV^{\pm}(\bbR)$ belong to $L^{2}(\bbR)$, irrespective of
the parameter values of $b_{1}$ and $z_{0}$. Hence, it manifests unbroken
SUSY. So, in this case position-dependent mass affects the symmetry breaking
scenario. But the mass profile $m(q)=\sech^{2}aq$, $q\in(-\infty,\infty)$
has no effect on dynamical breaking of $\cN$-fold SUSY which can be observed
by considering the leading behavior of the solvable sectors (\ref{e62}) and
(\ref{e63}). We have found that same is true for many other mass functions.

Also associated to this mass profile, one of the partner potentials given
in equation (\ref{e82}) is identical with the $V_{eff}(q)$ whose bound
state wave functions are given by exceptional $X_{1}$ Jacobi polynomials
[e.g., Eq.~(18) of Ref.~\cite{MR09a}], for the choice of parameters
$b_{1}=(\alpha-\beta+\cN)a^{2}/2$, $z_{0}=(\alpha+\beta)/(\alpha-\beta)$.
The simplified form of the other partner potential $U^{-}(q)$ matches
with the potential previously obtained in \cite{BGQR05} corresponding
to classical Jacobi polynomials. It is worth mentioning that this mass
profile $m(q)=\sech^{2}aq$ has been previously used in PDM Hamiltonians
of BenDaniel--Duke~\cite{BD66} and Zhu--Kroemer~\cite{ZK83} type
and interesting connection was shown \cite{Ba07} between the discrete
eigenvalues of such Hamiltonians and the stationary 1-soliton and 2-soliton
solutions of the Korteweg-de Vries (KdV) equation.

For the latter choice of the mass function, the change of variable is given
by $u(q)=\tan^{-1}(\sinh aq)/a$ and corresponding pair of potentials
$U^{\pm}(q)$ read as
\begin{align}
U^{\pm}(q)=&\;\frac{[2b_{1}(z_{0}+1)-\cN a^{2}z_{0}\mp(\cN-2)a^{2}]
 [2b_{1}(z_{0}+1)-\cN a^{2}z_{0}\mp(\cN+2)a^{2}]}{32a^{2}}\rme^{2aq}\notag\\
&+\frac{[2b_{1}(z_{0}-1)-\cN a^{2}z_{0}\pm(\cN-2)a^{2}]
 [2b_{1}(z_{0}-1)-\cN a^{2}z_{0}\pm(\cN+2)a^{2}]}{32a^{2}}\rme^{-2aq}\notag\\
&+\frac{1\pm1}{2}\frac{a^{2}}{z_{0}+1}\left[1-\frac{2(z_{0}-2)}{z_{0}-1
 +(z_0+1)\rme^{2aq}}-\frac{4(z_{0}-1)}{(z_{0}-1+(z_{0}+1)\rme^{2aq})^{2}}
 \right] \mp\frac{\cN b_{1}}{4} + V_0.
\label{e82}
\end{align}

\noindent
\textbf{Example 3.3.} $A(z)=(z-z_{0})^{2}/2$\\
Potentials:
\begin{subequations}
\label{e90}
\begin{align}
V^{(0)-}(q)&=\frac{(2b_{1}+\cN)^{2}z_{0}^{\,2}}{8}\rme^{-2q}
 +\frac{(4b_{1}^{\,2}-\cN^{\,2})z_{0}}{4}\rme^{-q}+V_{0},\\
V^{(0)+}(q)&=\frac{(2b_{1}+\cN)^{2}z_{0}^{\,2}}{8}\rme^{-2q}
 +\frac{(2b_{1}+\cN)^{2}z_{0}}{4}\rme^{-q}
 -\frac{z_{0}\rme^{-q}}{(1+z_{0}\rme^{-q})^{2}}+V_{0},
\end{align}
\end{subequations}
where $V_{0}$ is an irrelevant constant given by
\begin{align*}
V_{0}=\frac{b_{1}^{\,2}}{2}+\frac{b_{1}}{\cN}+\frac{\cN^{\,2}+11}{24}-R.
\end{align*}
Solvable sectors:
\begin{subequations}
\label{e83}
\begin{align}
\cV^{(0)-}&=\braket{1,z(q),z(q)^{2},\ldots}\exp\left[
 -\frac{(2b_{1}+\cN)z_{0}}{2}\rme^{-q}-\frac{\cN-1-2b_{1}}{2}q\right],\\
\cV^{(0)+}&=\braket{1,z(q)^{2},z(q)^{3},\ldots}z(q)^{-1}\exp\left[
 \frac{(2b_{1}+\cN)z_{0}}{2}\rme^{-q}-\frac{\cN-1+2b_{1}}{2}q\right].
\end{align}
\end{subequations}
We assume $z_{0}>0$ so that the pair of potentials $V^{(0)\pm}(q)$ has no
singularities in $(-\infty,\infty)$. As we will show in what follows,
the $\cN$-fold SUSY in this case can be partially broken. To see this,
we first introduce a pair of $k$-dimensional subspaces $\cV_{k}^{(0)\pm}$
of the solvable sectors $\cV^{(0)\pm}$ as
\begin{align}
\cV_{k}^{(0)-}&=\braket{1,z(q),\dots,z(q)^{k-1}}\exp\left[
 -\frac{(2b_{1}+\cN)z_{0}}{2}\rme^{-q}-\frac{\cN-1-2b_{1}}{2}q\right],\\
\cV_{k}^{(0)+}&=\braket{1,z(q)^{2},\dots,z(q)^{k}}z(q)^{-1}\exp\left[
 \frac{(2b_{1}+\cN)z_{0}}{2}\rme^{-q}-\frac{\cN-1+2b_{1}}{2}q\right].
\end{align}
Then, for a fixed $k\in\bbN$, we have
\begin{align}
\cV_{k}^{(0)-}(\bbR)\subset L^{2}(\bbR)&\quad\Llra\quad
 -\cN<2b_{1}<\cN+1-2k,\\
\cV_{k}^{(0)+}(\bbR)\subset L^{2}(\bbR)&\quad\Llra\quad
 2k-\cN-1<2b_{1}<-\cN.
\label{eq:ncon2}
\end{align}
{}From these conditions, it is easy to observe that $\cV_{k}^{(0)-}(\bbR)
\subset L^{2}(\bbR)$ if and only if $-\cN/2<b_{1}<(\cN+1-2k)/2$ for a
$k\in\bbN$ satisfying $k<\cN+1/2$, while there is no $k\in\bbN$ which
satisfy the condition (\ref{eq:ncon2}) and thus $\cV^{(0)+}(\bbR)\not\subset
L^{2}(\bbR)$ $\forall b_{1}\in\bbR$. Hence, the $\cN$-fold SUSY in the
constant-mass background is partially broken if there is a positive integer
$k\leq\cN$ for which the parameter $b_{1}$ satisfies
\begin{align*}
-\frac{\cN}{2}<b_{1}<\frac{\cN+1-2k}{2},
\end{align*}
and fully broken otherwise.

The solvable sectors $\cV^{\pm}$ of the corresponding PDM Hamiltonians
$H^{\pm}$ are written as
\begin{subequations}
\label{e84}
\begin{align}
\cV^{-}=&\;\braket{1,z(u(q)),z(u(q))^{2},\ldots}m(q)^{1/4}\notag\\
&\times\exp\left[-\frac{(2b_{1}+\cN)z_{0}}{2}\rme^{-u(q)}
 -\frac{\cN-1-2b_{1}}{2}u(q)\right],\\
\cV^{+}=&\;\braket{1,z(u(q))^{2},z(u(q))^{3},\ldots}z(u(q))^{-1}
 m(q)^{1/4}\notag\\
&\times\exp\left[\frac{(2b_{1}+\cN)z_{0}}{2}\rme^{-u(q)}
 -\frac{\cN-1+2b_{1}}{2}u(q)\right].
\end{align}
\end{subequations}
and the potentials $U^{\pm}(q)$ can be obtained using Eqs.~(\ref{eq:PCT1}),
(\ref{e90}a) and (\ref{e90}b).
We have checked the normalizability of the solvable sectors (\ref{e84})
with the following two mass functions.

(i) $m(q)=(1-q^{2})^{-1}$, $q\in(-1,1)$ for which the change
of variable is $u(q)=\sin^{-1}q$. This mass profile has been used in
Refs.~\cite{MR09b,CRS07} while considering the effective-mass quantum
nonlinear oscillator. This mass function has effect on dynamical
symmetry breaking because it manifests broken SUSY [i.e., neither
$\cV^{-}$ nor $\cV^{+}$ belongs to $ L^{2}(-1,1)$], which is
clear from the following expressions of $\cV^{-}$ and $\cV^{+}$:
\begin{align}
\cV^{-}=&\;\braket{1,z(u(q)),z(u(q))^{2},\ldots}
 \frac{1}{(1-q^{2})^{1/4}}\notag\\
 &\times\exp\left[-\frac{(2b_{1}+\cN)z_{0}}{2}\rme^{-\sin^{-1}q}
 -\frac{\cN-1-2b_{1}}{2}\sin^{-1}q\right],\\
\cV^{+}=&\;\braket{1,z(u(q))^{2},z(u(q))^{3},\ldots}
 \frac{1}{(1-q^{2})^{1/4}(\rme^{\sin^{-1}q}+z_{0})}\notag\\
 &\times\exp\left[\frac{(2b_{1}+\cN)z_{0}}{2}\rme^{-\sin^{-1}q}
 -\frac{\cN-1+2b_{1}}{2}\sin^{-1}q\right].
\end{align}

(ii) $m(q)=2\rme^{-2q^{2}}/\pi$ for which the solvable sectors (\ref{e84})
reduce to
\begin{align}
\cV^{-}=&\;\braket{1,z(u(q)),z(u(q))^{2},\ldots}\notag\\
&\times\exp\left[-\frac{q^{2}}{4}-\frac{(2b_{1}+\cN)z_{0}}{2}\rme^{-\ef q}
 -\frac{\cN-1-2b_{1}}{2}\ef q\right],\\
\cV^{+}=&\;\braket{1,z(u(q))^{2},\dots,z(u(q))^{\cN}}z(u(q))^{-1}\notag\\
&\times\exp\left[-\frac{q^{2}}{4}+\frac{(2b_{1}+\cN)z_{0}}{2}
 \rme^{-\ef q}-\frac{\cN-1+2b_{1}}{2}\ef q\right].
\end{align}
{}From the above solvable sectors, we observe that both $\cV^{-}(\bbR)$
and $\cV^{+}(\bbR)$ belong to $L^{2}(\bbR)$, irrespective of the
parameter value $b_{1}$, which means unbroken $\cN$-fold SUSY. Hence,
the mass function $m(q)=2\rme^{-2q^{2}}/\pi$ affects dynamical breaking of
the $\cN$-fold SUSY.

Hence, comparing the normalizability conditions in both the constant and
position dependent mass cases, we conclude that both the mass functions
change the behaviours of symmetry breaking.

\subsection{Effects of PDM on Dynamical Symmetry Breaking of Type $X_{2}$
 $\cN$-fold SUSY}\label{ss3.2}

In this section, we examine three different models of type $X_{2}$ $\cN$-fold
SUSY characterized by different choices of the two parameters $a_{1}$ and
$a_{2}$; $a_{1}\neq0$ and $a_{2}=0$ for the first model, $a_{1}=0$ and
$a_{2}\neq0$ for the second, and $a_{1}a_{2}\neq0$ for the third. The first
two choices lead to the rational- and hyperbolic-type potential pairs already
shown in Ref.~\cite{Ta10a}, while the last choice to an exponential-type
potential pair which is new and has not been investigated in the literature.

\noindent
\textbf{Example 3.4.} $A(z)=2z$ $[a_{1}=2]$.\\
Potentials:
\begin{subequations}
\label{e91}
\begin{align}
V^{(0)-}(q)=&\;\frac{q^{2}}{2}+\frac{4\alpha^{2}-1}{8q^{2}}
 +4\left[\frac{q^{2}-\alpha+1}{f(q^{2};\alpha)}
 -\frac{4(\alpha-1)q^{2}}{f(q^{2};\alpha)^{2}}\right]-\cN+V_{0},\\
V^{(0)+}(q)=&\;\frac{q^{2}}{2}+\frac{4(\alpha+\cN)^{2}-1}{8q^{2}}
 +4\left[\frac{q^{2}-\alpha-\cN+1}{f(q^{2};\alpha+\cN)}
 -\frac{4(\alpha+\cN-1)q^{2}}{f(q^{2};\alpha+\cN)^{2}}\right]+V_{0},
\end{align}
\end{subequations}
where $V_{0}=\cN-\alpha+3-c_{0}$ is an irrelevant constant.\\
Solvable sectors:
\begin{subequations}
\label{eqs:solv1}
\begin{align}
\cV_{\cN}^{(0)-}&=\braket{\tvph_{1}(q^{2};\alpha),\dots,\tvph_{\cN}
 (q^{2};\alpha)}\frac{q^{\alpha+1/2}\rme^{-q^{2}/2}}{f(q^{2};\alpha)},\\
\cV_{\cN}^{(0)+}&=\braket{\bar{\chi}_{1}(q^{2};\alpha+\cN),\dots,
 \bar{\chi}_{\cN}(q^{2};\alpha+\cN)}\frac{q^{-\alpha-\cN+1/2}
 \rme^{q^{2}/2}}{f(q^{2};\alpha+\cN)}.
\end{align}
\end{subequations}
In this case, the solvability condition $a_{2}(=a_{3}=a_{4})=0$ for type
$X_{2}$ is satisfied and thus the corresponding constant-mass Hamiltonians
$H^{(0)\pm}$ are simultaneously solvable.

For $\alpha>1$, a natural choice for the domain of these potentials is a
real half-line $S=\bbR_{+}$. On this domain $\bbR_{+}$, it is evident from
(\ref{eqs:solv1}) that $\cV_{\cN}^{(0)-}(\bbR_{+})\subset L^{2}(\bbR_{+})$
and $\cV_{\cN}^{(0)-}(\bbR_{+})\not\subset L^{2}(\bbR_{+})$. Therefore,
it manifests unbroken $\cN$-fold SUSY of the system in the constant-mass
background.

According to (\ref{eq:PCT3}), the solvable sectors $\cV_{\cN}^{\pm}$ of
the corresponding PDM Hamiltonians $H^{\pm}$ for an arbitrary mass function
$m(q)$ read as
\begin{align}
\cV_{\cN}^{-}&=\braket{\tvph_{1}(u(q)^{2};\alpha),\dots,\tvph_{\cN}
 (u(q)^{2};\alpha)}\frac{m(q)^{1/4}u(q)^{\alpha+1/2}
 \rme^{-u(q)^{2}/2}}{f(u(q)^{2};\alpha)},\\
\cV_{\cN}^{+}&=\braket{\bar{\chi}_{1}(u(q)^{2};\alpha+\cN),\dots,
 \bar{\chi}_{\cN}(u(q)^{2};\alpha+\cN)}\frac{m(q)^{1/4}
 u(q)^{-\alpha-\cN+1/2}\rme^{u(q)^{2}/2}}{f(u(q)^{2};\alpha+\cN)},
\end{align}
where $u(q)$ is given by (\ref{eq:uq}) and the PDM potentials $U^{\pm}(q)$
can be obtained using Eqs.~(\ref{eq:PCT1}) and (\ref{e91}).
In this case, we have not been able to find out any realistic mass function
which could break the $\cN$-fold SUSY. In other words, we can say that
the $\cN$-fold SUSY in this case is steady against many variations of mass
functions [e.g., $m(q)=\rme^{-q}$, $\sech^{2}q$].\\

\noindent
\textbf{Example 3.5.} $A(z)=(z^{2}+\zeta^{2})/2$, $[a_{2}=1/2$,
 $\zeta^{2}=(\alpha-1)(\alpha+\cN-1)>0]$.\\
Potentials:
\begin{align}
V^{(0)-}(q)=&\;\frac{\zeta^{2}}{8}\cosh^{2}q+\frac{\cN-1}{4}\zeta
 \sinh q+V_{0}\notag\\
&+\frac{1}{8\cosh^{2}q}\left[
 4(\cN-1)\zeta\sinh q+4\alpha^{2}+4(\cN-2)\alpha-\cN^{\,2}-2\cN
 +4\right]\notag\\
&-2(\alpha-1)\left[\frac{\zeta\sinh q-\alpha-\cN+3}{f(\zeta\sinh q;
 \alpha)}-2(\alpha-1)\frac{2\zeta\sinh q-\cN+1}{f(\zeta\sinh q;
 \alpha)^{2}}\right],
\label{e54}
\end{align}
\begin{align}
\lefteqn{
V^{(0)+}(q)=\frac{\zeta^{2}}{8}\cosh^{2}q+\frac{3\cN-1}{4}\zeta
 \sinh q+V_{0}}\notag\hspace{30pt}\\
&-\frac{1}{8\cosh^{2}q}\left[
 4(\cN+1)\zeta\sinh q-4\alpha^{2}-4(\cN-2)\alpha+\cN^{\,2}+6\cN
 -4\right]\notag\\
&-2(\alpha+\cN-1)\left[\frac{\zeta\sinh q-\alpha+3}{f(\zeta\sinh q;
 \alpha+\cN)}-2(\alpha+\cN-1)\frac{2\zeta\sinh q+\cN+1}{
 f(\zeta\sinh q;\alpha+\cN)^{2}}\right],
\label{e55}
\end{align}
where $V_{0}$ is an irrelevant constant given by
\begin{align*}
V_{0}=\frac{4\alpha^{2}+4(\cN-4)\alpha+\cN^{\,2}+16}{8}-c_{0}.
\end{align*}
Solvable sectors:
\begin{subequations}
\label{eqs:solv2}
\begin{align}
\cV_{\cN}^{(0)-}=&\;\braket{\tvph_{1}(\zeta\sinh q;\alpha),\dots,
 \tvph_{\cN}(\zeta\sinh q;\alpha)}\frac{\rme^{-\zeta(\sinh q)/2
 -\zeta \gd q}}{(\cosh q)^{\cN/2-1}f(\zeta\sinh q;\alpha)},\\
\cV_{\cN}^{(0)+}=&\;\braket{\bar{\chi}_{1}(\zeta\sinh q;\alpha+\cN),\dots,
 \bar{\chi}_{\cN}(\zeta\sinh q;\alpha+\cN)}\notag\\
&\times\frac{\rme^{\zeta(\sinh q)/2
 +\zeta\gd q}}{(\cosh q)^{\cN/2}f(\zeta\sinh q;\alpha+\cN)},
\end{align}
\end{subequations}
where $\gd q=\tan^{-1}(\sinh q)$ is the Gudermann function. The solvability
condition is not satisfied in this case and both of the Hamiltonians are
only quasi-solvable. For $\alpha>1$, the potentials $V^{\pm}(q)$ given
in (\ref{e54}) are defined on the whole real line $\bbR$. From
the solvable sectors (\ref{eqs:solv2}), it is clear that neither
$\cV_{\cN}^{(0)-}(\bbR)$ nor $\cV_{\cN}^{(0)+}(\bbR)$ belongs to
$L^{2}(\bbR)$, so the $\cN$-fold SUSY is dynamically broken in the
constant-mass background.

Now, the PDM potentials $U^{\pm}(q)$ can be obtained with help of
Eqs.~(\ref{eq:PCT1}), (\ref{e54}), and (\ref{e55}), and the solvable
sectors $\cV_{\cN}^{\pm}$ of the corresponding PDM Hamiltonians $H^{\pm}$
for an arbitrary mass function $m(q)$ read from (\ref{eq:PCT3}) as
\begin{subequations}
\label{e73}
\begin{align}
\cV_{\cN}^{-}=&\;\braket{\tvph_{1}(\zeta\sinh u(q);\alpha),\dots,
 \tvph_{\cN}(\zeta\sinh u(q);\alpha)}\notag\\
&\times\frac{m(q)^{1/4}\rme^{-\zeta(\sinh u(q))/2
 -\zeta \gd u(q)}}{(\cosh u(q))^{\cN/2-1}f(\zeta\sinh u(q);\alpha)},\\
\cV_{\cN}^{+}=&\;\braket{\bar{\chi}_{1}(\zeta\sinh u(q);\alpha+\cN),\dots,
 \bar{\chi}_{\cN}(\zeta\sinh u(q);\alpha+\cN)}\notag\\
&\times\frac{m(q)^{1/4}\rme^{\zeta(\sinh u(q))/2
 +\zeta\gd u(q)}}{(\cosh u(q))^{\cN/2}f(\zeta\sinh u(q);\alpha+\cN)},
\end{align}
\end{subequations}
where $u(q)$ is given by (\ref{eq:uq}). Let us now consider two cases:

(i) $m(q)=\sech^{2}q$, $q\in(-\infty,\infty),$
for which the change of variable is $u(q)=\gd q$. Then, the solvable
sectors of $U^{\pm}(q)$ are given by
\begin{subequations}
\label{e71}
\begin{align}
\cV_{\cN}^{-}=&\;\braket{\tvph_{1}(\zeta\sinh u(q);\alpha),\dots,
 \tvph_{\cN}(\zeta\sinh u(q);\alpha)}\notag\\
&\times\frac{\sqrt{\sech q}\,\rme^{-\zeta\sinh(\gd q)/2
 -\zeta\gd(\gd q)}}{[\cosh(\gd q)]^{\cN/2-1}f(\zeta\sinh u(q);\alpha)},\\
\cV_{\cN}^{+}=&\;\braket{\bar{\chi}_{1}(\zeta\sinh u(q);\alpha
 +\cN),\dots,\bar{\chi}_{\cN}(\zeta\sinh u(q);\alpha+\cN)}\notag\\
&\times\frac{\sqrt{\sech q}\,\rme^{\zeta\sinh(\gd q)/2
+\zeta\gd(\gd q)}}{[\cosh(\gd q)]^{\cN/2}f(\zeta\sinh u(q);\alpha+\cN)}.
\end{align}
\end{subequations}
In this case, the mass function as well as the potentials $U^{\pm}(q)$ are
well behaved on $(-\infty,\infty)$, so we can consider the whole real line
$\bbR$ as a domain of the potentials. From the solvable sectors (\ref{e71}),
it is clear that both $\cV_{\cN}^{\pm}(\bbR)$ belong to $L^{2}(\bbR)$,
which means unbroken $\cN$-fold SUSY, i.e., the mass profile affects
symmetry restoration.

(ii) $m(q)=2\rme^{-2q^{2}}/\pi$.
In this case, the solvable sectors $\cV_{\cN}^{\pm}$ reduce to
\begin{subequations}
\label{e74}
\begin{align}
\cV_{\cN}^{-}=&\;\braket{\tvph_{1}(\zeta\sinh u(q);\alpha),\dots,
 \tvph_{\cN}(\zeta\sinh u(q);\alpha)}\notag\\
&\times\frac{\exp[-q^{2}/4-\zeta\sinh(\ef q)/2-\zeta\gd(\ef q)]
 }{[\cosh(\ef q)]^{\cN/2-1}f(\zeta\sinh u(q);\alpha)},\\
\cV_{\cN}^{+}=&\;\braket{\bar{\chi}_{1}(\zeta\sinh u(q);\alpha+\cN),\dots,
 \bar{\chi}_{\cN}(\zeta\sinh u(q);\alpha+\cN)}\notag\\
&\times\frac{\exp[-q^{2}/4+\zeta\sinh(\ef q)/2+\zeta\gd(\ef q)]
 }{[\cosh(\ef q)]^{\cN/2}f(\zeta\sinh u(q);\alpha+\cN)}.
\end{align}
\end{subequations}
{}From the above solvable sectors $(\ref{e74})$, it is clear that both
$\cV_{\cN}^{\pm}(\bbR)$ belong to $L^{2}(\bbR)$. That is, in this case we
again have unbroken $\cN$-fold SUSY.

We note that there are other mass functions, e.g., $m(q)=(\beta+q^{2})^{2}/
(1+q^{2})^{2}$, which have no effect on the dynamical breaking of $\cN$-fold
SUSY, i.e., it is also possible to construct PDM systems which maintain
the broken $\cN$-fold SUSY.\\

\noindent
\textbf{Example 3.6.} $A(z)=(z+\zeta)^{2}/2$, $[a_{2}=1/2,
 a_{1}=\zeta=\sqrt{(\alpha-1)(\alpha+\cN-1)}]$.\\
Potentials:
\begin{align}
V^{(0)-}(q)=&\;\frac{1}{8}\rme^{2q}-\frac{\cN+1}{4}\rme^{q}
 -\frac{(\cN-1)(\cN+2\alpha-2\zeta-1)\zeta}{4}\rme^{-q}\notag\\
&+\frac{\zeta^{2}[\cN^{\,2}+2\cN(4\alpha-2\zeta-3)
 +4\alpha(2\alpha-2\zeta-3)+4\zeta+5]}{8}\rme^{-2q}\notag\\
&-2\left[\frac{(\alpha-\zeta-1)\rme^{q}}{f(\rme^{q}-\zeta;\alpha)}
 +\frac{2(\alpha-1)\rme^{2q}}{f(\rme^{q}-\zeta;\alpha)^{2}}\right]+V_{0},
\label{e46}
\end{align}
\begin{align}
V^{(0)+}(q)=&\;\frac{1}{8}\rme^{2q}+\frac{\cN-1}{4}\rme^{q}
 +\frac{(\cN+1)(\cN+2\alpha-2\zeta-1)\zeta}{4}\rme^{-q}\notag\\
&+\frac{\zeta^{2}[\cN^{\,2}+2\cN(4\alpha-2\zeta-3)
 +4\alpha(2\alpha-2\zeta-3)+4\zeta+5]}{8}\rme^{-2q}\notag\\
&-2\left[\frac{(\alpha+\cN-\zeta-1)\rme^{q}}{f(\rme^{q}-\zeta;\alpha
 +\cN)}+\frac{2(\alpha+\cN-1)\rme^{2q}}{f(\rme^{q}-\zeta;\alpha
 +\cN)^{2}}\right]+V_{0},
\label{e47}
\end{align}
where $V_{0}$ is an irrelevant constant given by
\begin{align*}
V_{0}=\frac{(\cN+2\alpha)^{2}+2\zeta(\cN-2\alpha)+2(7\zeta-8\alpha+8)}{8}
 -c_{0}.
\end{align*}
Solvable sectors:
\begin{align}
\cV_{\cN}^{(0)-}=&\;\frac{\left\langle \tvph_{1}(\rme^{q}-\zeta;\alpha),
 \dots,\tvph_{\cN}(\rme^{q}-\zeta;\alpha)\right\rangle}{f(\rme^{q}-\zeta;
 \alpha)}\notag\\
&\times\exp\left[-\frac{\rme^{q}}{2}+\frac{2\zeta-2\alpha-\cN+1}{2}\zeta
 \rme^{-q}-\frac{\cN-2}{2}q\right],
\label{e48}
\end{align}
\begin{align}
\cV_{\cN}^{(0)+}=&\;\frac{\left\langle\bar{\chi}_{1}(\rme^{q}-\zeta;\alpha
 +\cN),\dots,\bar{\chi}_{\cN}(\rme^{q}-\zeta;\alpha+\cN)\right\rangle}{
 f(\rme^{q}-\zeta;\alpha+\cN)}\notag\\
&\times\exp\left[\frac{\rme^{q}}{2}-\frac{2\zeta-2\alpha-\cN+1}{2}\zeta
 \rme^{-q}-\frac{\cN}{2}q\right].
\label{e49}
\end{align}
This system is new and presented in this paper for the first time.
The exponential-type $V_{\cN}^{\pm}(q)$ are naturally defined on the whole
real line $\bbR$ since they have no singularity on it, so the Hilbert
space is $L^{2}(\bbR)$. Noting that $2\zeta-2\alpha-\cN+1<0$ for $\alpha>1$,
since
\begin{align*}
4\zeta^{2}-(2\alpha+\cN-1)^{2}=-4\alpha-(\cN-1)(\cN+3)<-(\cN+1)^{2}<0,
\end{align*}
we see that $\cV_{\cN}^{(0)-}(\bbR)\subset L^{2}(\bbR)$ and
$\cV_{\cN}^{(0)+}(\bbR)\not\subset L^{2}(\bbR)$ for $\zeta>0$. Hence, it
manifests unbroken $\cN$-fold SUSY. For $\zeta<0$, on the other hand,
neither $\cV_{\cN}^{(0)-}(\bbR)$ nor $\cV_{\cN}^{(0)+}(\bbR)$ belongs to
$L^{2}(\bbR)$, so the $\cN$-fold SUSY is broken in the constant-mass
background.

In a PDM background, the solvable sectors $\cV_{\cN}^{\pm}$ of the type
$X_{2}$ PDM Hamiltonians $H^{\pm}$ are deformed as [cf., Eq.~(\ref{eq:PCT3})]
\begin{align}
\cV_{\cN}^{-}=&\;\frac{\braket{\tvph_{1}(\rme^{u(q)}-\zeta;\alpha),
 \dots,\tvph_{\cN}(\rme^{u(q)}-\zeta;\alpha)}}{f(\rme^{u(q)}
 -\zeta;\alpha)}\notag\\
&\times m(q)^{1/4}\exp\left[-\frac{e^{u(q)}}{2}+\frac{2\zeta-2\alpha
 -\cN+1}{2}\zeta\rme^{-u(q)}-\frac{\cN-2}{2}u(q)\right],
\label{e75}
\end{align}
\begin{align}
\cV_{\cN}^{+}=&\;\frac{\braket{\bar{\chi}_{1}(\rme^{u(q)}-\zeta;
 \alpha+\cN),\dots,\bar{\chi}_{\cN}(\rme^{u(q)}-\zeta;\alpha+\cN)}
 }{f(\rme^{u(q)}-\zeta;\alpha+\cN)}\notag\\
&\times m(q)^{1/4}\exp\left[\frac{\rme^{u(q)}}{2}-\frac{2\zeta-2\alpha
 -\cN+1}{2}\zeta\rme^{-u(q)}-\frac{\cN}{2}u(q)\right],
\label{e76}
\end{align}
and the potentials $U^{\pm}(q)$ can be obtained using Eqs.~(\ref{eq:PCT1}),
(\ref{e46}), and (\ref{e47}). In this case, the choice of mass functions
are as follows:

(i) $m(q)=(1-q^{2})^{-1}$, $q\in(-1,1)$, for which the solvable sectors of
the PDM Hamiltonians $H^{\pm}$ are given by
\begin{align}
\cV_{\cN}^{-}=&\;\frac{\braket{\tvph_{1}(\rme^{u(q)}-\zeta;\alpha),
 \dots,\tvph_{\cN}(\rme^{u(q)}-\zeta;\alpha)}}{(1-q^{2})^{1/4}
 f(\rme^{u(q)}-\zeta;\alpha)}\notag\\
&\times\exp\left[-\frac{\rme^{\sin^{-1}q}}{2}+\frac{2\zeta-2\alpha-\cN+1}{2}
 \zeta\rme^{-\sin^{-1}q}-\frac{\cN-2}{2}\sin^{-1}q\right],
\label{e78}
\end{align}
\begin{align}
\cV_{\cN}^{+}=&\;\frac{\braket{\bar{\chi}_{1}(\rme^{u(q)}-\zeta;\alpha
 +\cN),\dots,\bar{\chi}_{\cN}(\rme^{u(q)}-\zeta;\alpha+\cN)}}{
 (1-q^{2})^{1/4}f(\rme^{u(q)}-\zeta;\alpha+\cN)}\notag\\
&\times\exp\left[\frac{\rme^{\sin^{-1}q}}{2}-\frac{2\zeta-2\alpha-\cN+1}{2}
 \zeta\rme^{-\sin^{-1}q}-\frac{\cN}{2}\sin^{-1}q\right].
\label{e79}
\end{align}
{}From the above solvable sectors, it is clear that both $\cV_{\cN}^{\pm}
(-1,1)$ do not belong to $L^{2}(-1,1)$, so it manifests broken $\cN$-fold
SUSY irrespective of the sign of $\zeta$. Hence, comparing the normalizability
conditions in both the constant and position-dependent mass cases, we conclude
that the mass function $m(q)=(1-q^{2})^{-1}$ affects dynamical
breaking of $\cN$-fold SUSY for $\zeta>0$.

(ii) $m(q)=2\rme^{-2q^{2}}/\pi$, $q\in(-\infty,\infty)$, for which the
$\cN$-fold SUSY remains unbroken, which is evident from the corresponding
solvable sectors given by
\begin{align}
\cV_{\cN}^{-}=&\;\frac{\braket{\tvph_{1}(\rme^{u(q)}-\zeta;\alpha),
 \dots,\tvph_{\cN}(\rme^{u(q)}-\zeta;\alpha)}}{f(\rme^{u(q)}
 -\zeta;\alpha)}\exp\left[-\frac{q^{2}}{4}\right.\notag\\
&\left.-\frac{\rme^{\ef q}}{2}+\frac{2\zeta-2\alpha-\cN+1}{2}\zeta
 \rme^{-\ef q}-\frac{\cN-2}{2}\ef q\right],
\label{e80}
\end{align}
\begin{align}
\cV_{\cN}^{+}=&\;\frac{\braket{\bar{\chi}_{1}(\rme^{u(q)}-\zeta;
 \alpha+\cN),\dots,\bar{\chi}_{\cN}(\rme^{u(q)}-\zeta;\alpha+\cN)}
 }{f(\rme^{u(q)}-\zeta;\alpha+\cN)}\notag\\
&\times\exp\left[-\frac{q^{2}}{4}+\frac{\rme^{\ef q}}{2}
 -\frac{2\zeta-2\alpha-\cN+1}{2}\zeta\rme^{-\ef q}-\frac{\cN}{2}\ef q\right].
\label{e81}
\end{align}
{}From the normalizability conditions in the constant and position-dependent
mass cases, we see that the mass function $m(q)=2\rme^{-2q^{2}}/\pi$ affects
the dynamical breaking of $\cN$-fold SUSY for $\zeta<0$.

\section{Summary and Perspectives}
\label{s5}

In this paper, we have investigated effect of position-dependent mass
background on dynamical breaking of type B and type $X_{2}$ $\cN$-fold SUSY.
We have selected three different models in the constant mass background for
each type, and then examined whether some of the physically relevant effective
mass profiles can affect the pattern of $\cN$-fold SUSY breaking in each
model. We summarize the results in Table~\ref{tb:result}. We can easily see from
Table~\ref{tb:result} that, except for the rational potentials, some of the PDM
profiles can actually affect and change the patterns of dynamical $\cN$-fold
SUSY breaking in all the trigonometric, hyperbolic, and exponential potentials.
Although we have selected the specific types of $\cN$-fold SUSY to develop
physical applicability of the new mathematical concept of exceptional polynomial
subspaces, we can of course make a similar analysis on other types of $\cN$-fold
SUSY such as type A and type C to find out positive effect of PDM on SUSY
breaking in some models.

Hence, it would be possible to observe experimentally transition between a broken
and an unbroken phases if an effective mass can be controlled experimentally
such that the constant mass limit can be also realized in an experimental setting.
The physical meanings of a position-dependent mass depend on each physical
system under consideration, for instance, the curvature of the local band
structure of an alloy in the momentum space for electrons in a crystal with
graded composition~\cite{GK93}, the particle densities of 3He and 4He
in pure and mixed helium clusters with doping atoms or molecules~\cite{BPGHN97},
the effective electron mass for electrons confined in a quantum dot~\cite{SL97}
and for dipole excitations of sodium clusters~\cite{PSC94}, and so on.
Thus, if we can prepare such an atomic, molecular, or condensed matter system
which is described by a certain PDM quantum model subjected to an $\cN$-fold
SUSY potential with mass profiles, e.g., $m(q)=\rme^{-\nu^{2}q^{2}}$
or $(1-\nu^{2}q^{2})^{-1}$ where $\nu$ is an experimentally adjustable parameter
such that $\nu\to0$ is realizable, then the spectral change of the system could
be observed at $\nu=0$ due to the phase transition. The essence and novelty of
our idea rely on the observation that the physically controllable PDM can
cause the phase transition by changing the normalizability of the solvable
sector although the latter is superficially a simple mathematical aspect. Hence,
it is quite important to note that the normalizability of wave functions can play
much more roles than the quantization of energy spectrum which is referred to
by any standard textbook on quantum mechanics.

\begin{table}
\caption{The effects of PDM profiles on dynamical breaking of $\cN$-fold SUSY
 in various type B and type $X_{2}$ models.}
\begin{tabular}{|c|c|c|c|}
 \hline
 \multicolumn{2}{|c}{Types of potentials } & \multicolumn{2}{|c|}{Dynamical
  breaking of $\cN$-fold SUSY}\\
 \cline{3-4}
 \multicolumn{2}{|c|}{} & Constant mass & PDM\\
 \hline
  & rational & unbroken  & no effect\\
 \cline{2-4}
 Type B &  trigonometric & broken & unbroken for $m(q)\propto\rme^{-2q^{2}}$\\
 \cline{2-4}
  && partially broken for & broken for $m(q)=(1-q^{2})^{-1}$\\
  & exponential & $-\cN/2<b_{1}<(\cN+1-2k)/2$ &\\
  && and fully broken otherwise   & unbroken for $m(q)\propto\rme^{-2q^{2}}$\\
 \hline
  & rational & unbroken  & no effect\\
 \cline{2-4}
 Type $X_{2}$& hyperbolic & broken  & unbroken for $m(q)\propto\sech^{2}q$,
  $\rme^{-2q^{2}}$\\
 \cline{2-4}
  & exponential & unbroken for $\zeta>0$   & broken for $m(q)=(1-q^{2})^{-1}$,
   $\forall\zeta$\\
  && broken for $\zeta<0$ & unbroken for $m(q)\propto\rme^{-2q^{2}}$,
   $\forall\zeta$\\
 \hline
\end{tabular}
\label{tb:result}
\end{table}

We note that this experimental observability might have impact not only on some
atomic, molecular, and condensed matter problems from which PDM quantum
theory originated, but also
on high-energy physics. Until now many high-energy physicists have believed
that SUSY is realized at the GUT or Planck scale as a resolution of the
naturalness and the hierarchy problem but is broken at least at the
electroweak scale. Unfortunately, however, theoretical analysis on dynamical
SUSY breaking in field theoretical models are extremely difficult on the one
hand, and it is virtually impossible to make a GUT scale experiment on the
other hand. The aforementioned experimental observability suggests that
we might extract some clues to understand dynamical SUSY breaking in
high-energy physics from realistic eV scale experiments in atomic, molecular,
and condensed matter
physics. It is because the Witten's work~\cite{Wi81} has indicated
that the mechanism of dynamical SUSY breaking in quantum field theory and
quantum mechanics is essentially the same. We also note that the careful
non-perturbative analyses in Refs.~\cite{AKOSW99,ST02} have shown that
the mechanism of dynamical breaking of ordinary and $\cN$-fold SUSY is
also the same. Hence, dynamical aspects of SUSY quantum field theoretical
models would be mimicked in $\cN$-fold SUSY quantum mechanical toy
models, regardless of whether or not $\cN$-fold SUSY can be realized in higher
dimensions. Therefore, we believe that further studies in this direction are
worth pursuing both theoretically and experimentally. From a theoretical point
of view it is a challenging issue to investigate both a perturbation theory and
the non-renormalization theorem in PDM quantum systems.

\begin{acknowledgments}

This work (T.~T.) was partially supported by the National Cheng Kung
University under the grant No.\ HUA:98-03-02-227.

\end{acknowledgments}



\bibliography{refsels}
\bibliographystyle{npb}

\end{document}